\documentclass{article}

\usepackage{arxiv}

\usepackage[utf8]{inputenc} 
\usepackage[T1]{fontenc}    
\usepackage{hyperref}       
\usepackage{url}            
\usepackage{booktabs}       
\usepackage{amsfonts}       
\usepackage{nicefrac}       
\usepackage{microtype}      
\usepackage{lipsum}		
\usepackage{graphicx}
\usepackage{doi}

\usepackage{rotating,tabularx}
\usepackage{caption}
\usepackage{array} 
\usepackage[numbers,sort&compress]{natbib}
\title{Publicly available datasets of breast histopathology H\&E whole-slide images: A scoping review}

\author{
    \begin{tabular}{c}
        \href{https://orcid.org/0000-0000-0000-0000}{\includegraphics[scale=0.06]{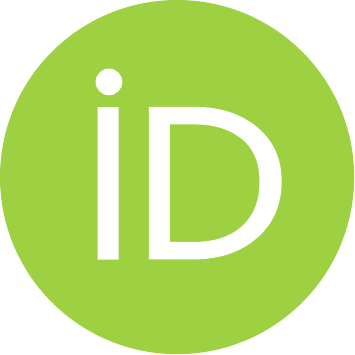}\hspace{1mm}Masoud Tafavvoghi} \\
        Department of Community Medicine \\
        UiT The Arctic University of Norway \\
        \texttt{masoud.tafavvoghi@uit.no} \\
    \end{tabular}
    \vspace{5mm} 
    \and
    \begin{tabular}{c}
        Lars Ailo Bongo \\
        Department of Computer Science \\
        UiT The Arctic University of Norway \\
    \end{tabular}
    \vspace{5mm} 
    \and
    \begin{tabular}{c}
        Nikita Shvetsov \\
        Department of Computer Science \\
        UiT The Arctic University of Norway \\
    \end{tabular}
    \vspace{5mm} 
    \and
    \begin{tabular}{c}
        Lill-Tove Rasmussen Busund \\
        Department of Medical Biology \\
        UiT The Arctic University of Norway \\
    \end{tabular}
    \vspace{5mm} 
    \and
    \begin{tabular}{c}
        Kajsa Møllersen \\
        Department of Community Medicine \\
        UiT The Arctic University of Norway \\
    \end{tabular}
}

\date{}

\renewcommand{\shorttitle}{\textit{ }}

\hypersetup{
pdftitle={Publicly available datasets of breast histopathology H\&E whole-slide images: A scoping review},
pdfsubject={q-bio.NC, q-bio.QM},
pdfauthor={David S.~Hippocampus, Elias D.~Striatum},
pdfkeywords={Breast cancer, Deep learning, Computational pathology, Publicly available dataset, Whole slide image},
}

\begin{document}
\maketitle

\begin{abstract}
Advancements in digital pathology and computing resources have made a significant impact in the field of computational pathology for breast cancer diagnosis and treatment. However, access to high-quality labeled histopathological images of breast cancer is a big challenge that limits the development of accurate and robust deep learning models. In this scoping review, we identified the publicly available datasets of breast H\&E stained whole-slide images (WSI) that can be used to develop deep learning algorithms. We systematically searched nine scientific literature databases and nine research data repositories and found 17 publicly available datasets containing 10385 H\&E WSIs of breast cancer. Moreover, we reported image metadata and characteristics for each dataset to assist researchers in selecting proper datasets for specific tasks in breast cancer computational pathology. In addition, we compiled two lists of breast H\&E patches and private datasets as supplementary resources for researchers. Notably, only 28\% of the included articles utilized multiple datasets, and only 14\% used an external validation set, suggesting that the performance of other developed models may be susceptible to overestimation. The TCGA-BRCA was used in 52\% of the selected studies. This dataset has a considerable selection bias that can impact the robustness and generalizability of the trained algorithms. There is also a lack of consistent metadata reporting of breast WSI datasets that can be an issue in developing accurate deep learning models, indicating the necessity of establishing explicit guidelines for documenting breast WSI dataset characteristics and metadata.
\end{abstract}


\section{Introduction}
\begin{figure*}
	\centering
	\includegraphics[scale=.49]{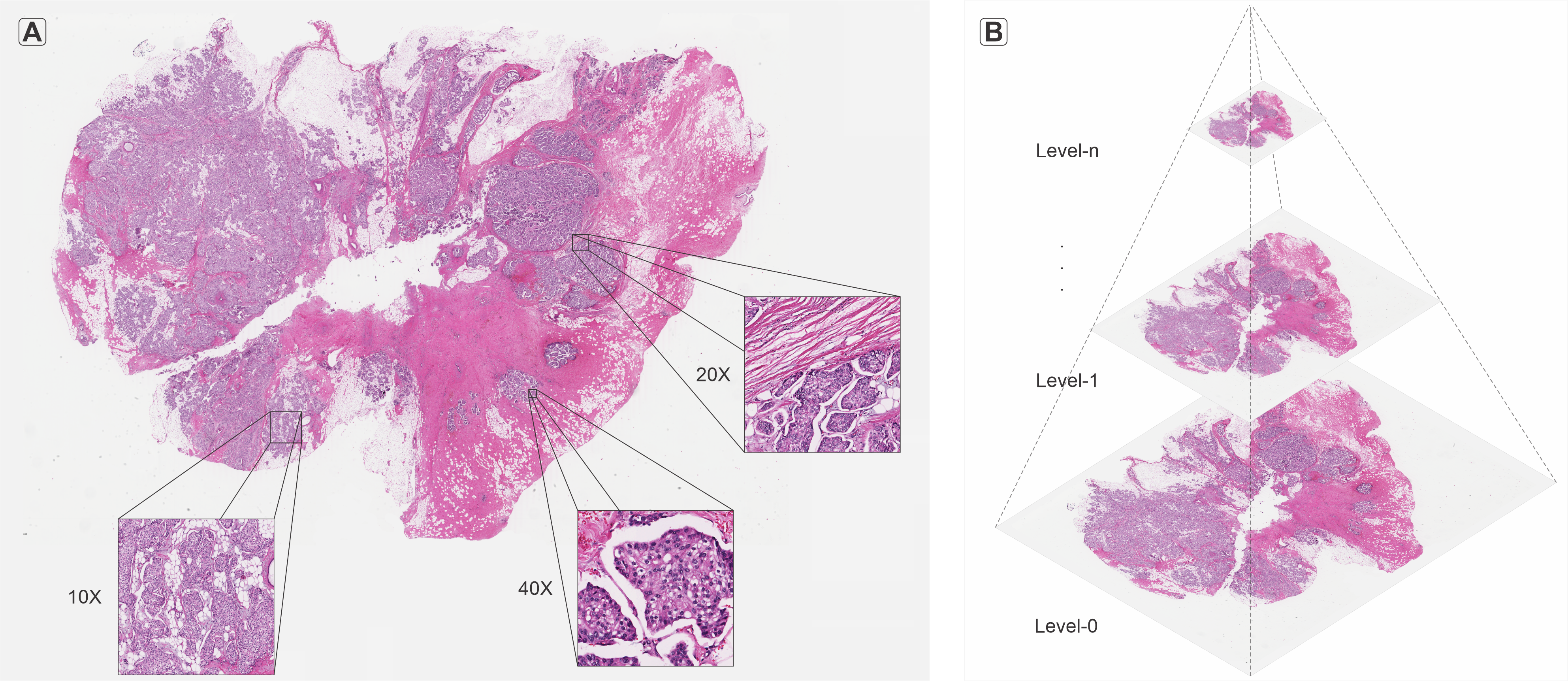}
	\caption{An example of breast H\&E WSI from TCGA-BRCA dataset: (A) illustration of 3 tile images in 10$\times$, 20$\times$, and 40$\times$ magnifications in a WSI; (B) WSI's multi-resolution pyramid with the highest resolution in level-0.}
	\label{FIG:1}
\end{figure*}

One important area of active research in pathology is the use of deep learning for analyzing H\&E histopathology whole-slide images (WSI) - the gold standard for the clinical diagnosis of cancer~\cite{liu2022classification}. Deep learning algorithms can identify complex patterns in billion-pixel microscope images that may not be readily apparent to human experts (an example is shown in Fig.~\ref{FIG:1}). For instance, deep learning models have been used to predict breast cancer recurrence~\cite{yang2022prediction} and classify breast cancer subtypes~\cite{srikantamurthy2023classification} by using histopathological WSIs.

The lack of adequately labeled datasets is a significant limitation in computational pathology for breast cancer, as the performance of deep learning algorithms depends on the availability of sufficient high-quality training and validation data. The use of large and diverse training datasets allows algorithms to identify complex patterns and nonlinear relationships more accurately. In addition, using large and independent validation datasets can increase the reliability of models and mitigate overfitting risk, which in turn improves the generalizability of the models~\cite{hinton2012improving}.

With advances in technology to produce and utilize data, there is a growing recognition of the benefits of data sharing, so greater openness in scientific research is being advocated for by scientists. The FAIR principles~\cite{wilkinson2016fair}, developed in response to this push for open-access data, provide a framework for making research data more Findable, Accessible, Interoperable, and Reusable. Based on the FAIR principles, medical data should be easily findable by both humans and machines by using a standardized and well-documented approach for metadata and data description and by making use of appropriate metadata standards, taxonomies, and ontologies. Data should be made accessible to all researchers authorized to access it per relevant ethical and legal frameworks. It should be feasible to integrate the data with other datasets and software tools in a seamless manner. Interoperability can be facilitated by using open data formats, data models, and data dictionaries. In addition, data should be designed with the intention of supporting data reuse, allowing other researchers to build upon the data and reproduce the results. To support data reuse, data should be accompanied by complete and accurate documentation, licensing information, and data citation. This approach enriches access to larger and more diverse datasets~\cite{gargeya2017automated}, enhances faster development of deep learning models~\cite{ching2018opportunities}, and improves their accuracy and performance~\cite{rajkomar2019machine}, which can consequently lead to better patient outcomes and improved quality of care~\cite{shickel2017deep}.

Like other medical data, histopathology images are protected by ethical and legal regulations related to privacy, security, and consent. To share medical data, data owners should adhere to relevant regulatory requirements, such as the General Data Protection Regulation (GDPR) in the European Union and the Health Insurance Portability and Accountability Act (HIPAA) in the United States. After the EU GDPR revised the new regulations in May 2018~\cite{simell2019transnational}, medical data sharing has become more challenging due to concerns over maintaining the confidentiality of patient health information, especially in light of high-profile data breaches and incidents of data misuse. In addition to privacy and security concerns, intellectual property and regulatory issues can pose significant barriers to sharing medical data. Resource constraints can also limit the ability of researchers to share medical data. The costs associated with collecting, storing, and curating medical data can be substantial, and many researchers may lack the necessary resources or expertise to manage and share their data effectively.

As the field of big data analysis continues to expand, having access to summaries of existing data has become more advantageous for researchers. Such overviews can assist in identifying relevant datasets without having to begin a new data collection process. Additionally, having a comprehensive and standardized set of public datasets can facilitate the reproducibility and comparability of research findings across different studies. A systematic review of available public datasets can help to identify gaps and limitations in the existing datasets and opportunities for improving the quality and diversity of available datasets. To address this, Hulsen~\cite{hulsen2019overview} has conducted a systematic review, providing an overview of publicly available patient-centered datasets of prostate cancer presented in imaging, clinical, and genomics categories. He identified 42 publicly available datasets that can efficiently support prostate cancer researchers in selecting appropriate data resources. He found that most datasets do not follow the FAIR principles, as some have legacy issues and need decoding work that might increase the possibility of human error.
In~\cite{wen2022characteristics}, the authors have systematically reviewed characteristics of publicly available datasets of skin cancer images, which can be leveraged for the advancement of machine learning algorithms for skin cancer diagnosis. They have reported 21 open-access datasets and 17 open-access atlases available for data extraction. They came to the conclusion that there is inconsistency in reporting image metadata, and population representations are limited in open-access datasets of skin cancer. Leung et al.~\cite{leung2015machine} have reviewed the datasets available for machine learning in genomic medicine, including an overview of available omic datasets. They suggested using multiple data sources to rectify problems arising from the missing information from individual datasets.

This scoping review aims to identify and assess the characteristics of all publicly available datasets of breast H\&E WSIs to reduce the demand for setting up new studies to collect data for the development of deep learning algorithms. This overview helps to identify potential data sources, the suitability of each dataset for specific tasks in computational pathology, and their quality and biases to ensure the generalizability of machine learning models. To the best of our knowledge, there has not been any study specifically targeting the available datasets of breast H\&E WSIs. However, several studies~\cite{brancati2022bracs, zeiser2021breast, duggento2021deep, hamidinekoo2018deep, liew2021review} have mentioned a small number of such datasets, suggesting the necessity for a comprehensive overview of all available datasets in this field.

\section*{Methods}

We conducted a scoping review based on the PRISMA-ScR guidelines~\cite{tricco2018prisma} to identify all publicly available breast H\&E WSI datasets appropriate for deep learning (\nameref{Supplement1}). Since this scoping review does not evaluate direct health outcomes, it was not eligible to be registered with PROSPERO~\cite{booth2012nuts}. 

Our inclusion criteria were papers using, reviewing, or mentioning any publicly available dataset of human breast H\&E WSIs. These may be introduced in machine learning challenges and contests or published for research purposes. A search was conducted in July 2023 using the following criteria: ("deep learning" OR "machine learning") AND ("whole slide images" OR WSI) AND (breast) AND (histology OR histopathology OR pathology) AND (data OR dataset OR "data set"). In total, nine scientific literature databases were queried: Pubmed, Medline, MDPI, Web of Science, Science Direct, Semantic Scholar, IEEE-Explore, Association for Computing Machinery (ACM) digital library, and the dbpl computer science bibliography. The search for our queries were not limited to only titles and abstracts but across all fields in the search engines, including full-text content and other relevant data fields. To ensure a manageable scope for this review, results were limited to full-text articles in English published between the years 2015-2023, and the following exclusion criteria were applied:
\begin{enumerate}
    \item not of human breast tissue, for instance, use of histology images of canine or mouse
    \item using other modalities like CT or MRI instead of histology images
    \item images of other organs (like lung, skin, etc.) rather than breast
    \item only patch or image tile datasets instead of whole-slide images
    \item non-image data like genomics or clinical data 
    \item tissues not stained with H\&E, e.g. immunohistochemistry (IHC) images
    \item not publicly available datasets
    \item use of unoriginal or subset datasets derived entirely from public datasets
\end{enumerate}

\begin{figure*}[!ht]
	\centering
	\includegraphics[scale=.75]{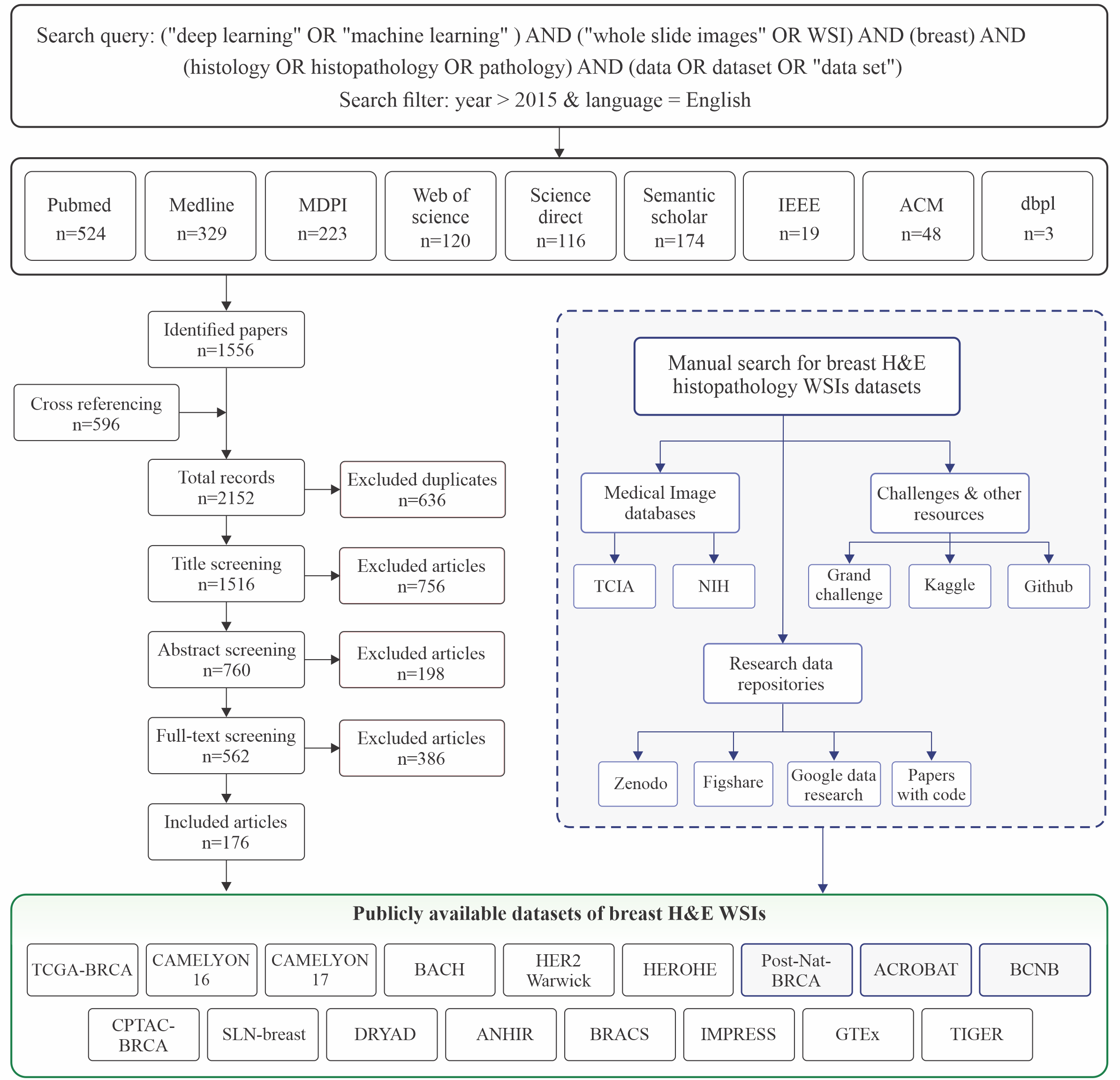}
	\caption{Data collection workflow. The dashed line box represents the manual search for publicly available datasets of breast H\&E WSIs.}
	\label{FIG:2}
\end{figure*}

 Fig.~\ref{FIG:2} summarizes the workflow diagram of the data collection. Of the 2152 articles from the search results and cross-referencing, 636 articles were removed as duplicates. The remaining 1516 identified articles were then screened by title, abstract, and full-text, respectively, by two independent reviewers (MT and (KM or LAB or NS)) by using the Mendeley reference management software. Subsequently, studies meeting the inclusion criteria were meticulously chosen and annotations were added to indicate the datasets utilized in each selected paper. In the event of disagreement on the inclusion, a third co-author (KM or LB) checked the article to make the final decision on whether it should be included or not. Out of 1516 articles, 756, 198, and 386 were excluded by their title, abstract and full-text assessment, respectively. This resulted in 176 articles that were included in this review. 

In addition to the scientific literature databases, we searched nine online databases and repositories known to contain public datasets. We used search strings in the~\nameref{Supplement2} to find breast histopathology datasets in The Cancer Imaging Archive, US National Institutes of Health (NIH), Google Data Research, Zenodo, Figshare, Github, Kaggle, Grand Challenge, and the Papers with Code platforms. All the relevant search results that featured breast histopathological images were reviewed by the first reviewer (MT), and their associated metadata and documentation were examined to find other public datasets of breast H\&E WSIs that were not included in the selected papers.

\section*{Results}

The 176 included papers used or reviewed one or more of 14 public datasets of breast H\&E WSIs (Table~\ref{t1}). In addition, we manually identified three datasets: the Post-Nat-BRCA dataset in the Cancer Imaging Archive repository and the ACROBAT and BCNB datasets in the Grand Challenge website, none of which were used in any of the selected articles. These 17 publicly available datasets comprise 10385 breast H\&E WSIs appropriate for machine learning use.
An additional 89 datasets of histologic breast imagery were identified but were not included as they did not fall within the scope of this review. Of these, 32 datasets were tiles (\nameref{Supplement3}) rather than WSIs, and 57 datasets were privately held or required data use agreements rather than being publicly available (\nameref{Supplement4}). The corresponding clinical data has been published for only six datasets. Interested readers can find detailed information on the available clinical variables for each dataset in \nameref{Supplement5}.

\begin{sidewaystable*}
    \centering
    \renewcommand{\arraystretch}{1.35} 
    \captionsetup{justification=raggedright,singlelinecheck=false} 
    \caption{Publicly available datasets of breast histological H\&E WSIs.}
    \label{t1}
    \small 
    \begin{tabular}{p{1.65cm}p{1.15cm}p{1.65cm}p{0.75cm}p{0.8cm}p{1.0cm}p{4.7cm}p{1.1cm}p{2.4cm}p{1.4cm}p{0.7cm}p{0.99cm}}
        \toprule
        {Dataset} & {AKA} & {Source} & {Pub. year} & {WSIs} & {Patients} & {Annotations/ Labels} & {Clinical data} & {Scanner} & {Pixel size ($\mu$m/pixel)} & {Size (GB)} & {Image format} \\
        \midrule
        ACROBAT & - & Sweden & 2023 & 1153 & 1153 & Landmark pairs for the WSIs in the validation and test sets  & - & NanoZoomer X360 and XR  & 0.91 & 1164 & TIFF \\
        ANHIR & - & Spain & 2019 & 5 & - & Coordinates of tumor area & - & Aperio AT2 & 0.25 & 0.2 & JPG \\
        BACH & ICIAR 2018 & Portugal & 2018 & 30 & - & 10 WSIs have coordinates of ROIs, labeled pixel-wise & - & Leica SCN400 & 0.50 & 7 & SVS, TIFF \\
        BCNB & - & China & 2022 & 1058 & 1058 & Coordinates of tumor regions & \checkmark & Iscan Coreo & - & 33 & JPG \\
        BRACS & - &  Italy & 2020 & 547 & 189 & Labels for 6 different subtypes & - & Aperio AT2 & 0.25 & 1100 & SVS \\
        Camelyon16 & - & Netherlands & 2016 & 399 & 399 & ROI polygons, pN-stage labels & - & Pannoramic 250, NanoZoomer-XR & 0.24 & 1160 & TIFF\\
        Camelyon17 & - & Netherlands & 2017 & 1000 & 200 & ROI polygons, pN-stage labels & - & Pannoramic 250, NanoZoomer-XR, Philips IntelliSite & 0.24  & 2950 & TIFF \\
        CPTAC-BRCA & CPTAC &  USA & 2021 & 642 & 134 & PAM50 molecular subtypes, tumor stage, etc. from the clinical data & \checkmark & - & 0.25, 0.50 & 113 & SVS \\
        DRYAD & - & USA & 2018 & 584 & - & ROI polygons, binary masks of invasive regions & - & - & - & 4.6 & PNG \\
        GTEx-breast & - &  USA & - & 894 & 894 & - & \checkmark & - & 0.50 & 80 & SVS \\
        HER2 & Warwick &  UK & 2016 & 86 & 86 & HER2 score, percentage of cells with complete membrane staining & - & NanoZoomer C9600 & 0.23 & 20 & SVS\\
        HEROHE & - &  Portugal & 2020 & 500 & 360 & Binary labels of HER2+/- status & - & Pannoramic 1000 & 0.24 & 820 & MRXS \\
        IMPRESS & - &  USA & 2023 & 126 & 126 & Histologic subtypes, tumor size, response to therapy, etc. & \checkmark & Hamamatsu & 0.50 & 27 & SVS \\
        Post-NAT-BRCA & - & Canada & 2021 & 96 & 54 & Cellularity and cell label, ER, PR, and HER2 scores & \checkmark & Aperio & 0.50 & 43 & SVS \\
        SLN- Breast & - &  USA & 2021 & 130 & 78 & Binary labels of metastasis status & - & Aperio & 0.50 & 53 & SVS \\
        TCGA- BRCA & TCGA &  USA & - & 3111 & 1098 & Tumor histology and molecular subtypes, given treatment, etc. from the clinical data & \checkmark & - & 0.25 & 1640 & SVS \\
        TIGER & - &  Netherlands & 2022 & 370 & 370 & ROI polygons, Lymphocyte and Plasma cells indicators, TIL values & - & - & 0.50 & 169 & TIFF \\
        \bottomrule
    \end{tabular}
\end{sidewaystable*}

There are also three public datasets of breast histopathological whole-slide images that have acquired all or part of the data from other publicly available datasets: TUPAC16~\cite{TUPAC16}, DRYAD, and TIGER with 821, 195, and 151 WSIs from the TCGA-BRCA, respectively (derived datasets of image tiles can be found in~\nameref{Supplement3}). Therefore, there is a risk of obtaining an overly optimistic performance estimate for trained models if derived datasets are used for the validation. This is because the model has already seen some of the data during training, and using derived datasets for validation may lead to an overestimation of the model's ability to generalize to new, unseen data. However, such derived datasets may be published with extra information not provided in the original datasets. For example, TUPAC16 has 500 WSIs in the training set, all derived from the TCGA-BRCA, with corresponding tumor proliferation and molecular proliferation scores as ground truth which were not included in the original TCGA-BRCA dataset. Including such extra information would be highly advantageous in developing models in breast computational pathology. Table~\ref{t2} shows details of the derived datasets of breast histopathology WSIs.

\begin{table*}[!ht]
    \centering
    \renewcommand{\arraystretch}{1.35} 
    \caption{Derived datasets of breast H\&E WSIs.}
    \label{t2}
    \small
    \begin{tabular}{p{1.3cm}p{0.6cm}p{0.7cm}p{1.0cm}p{4.8cm}p{5.4cm}}
        \toprule
        Dataset & Year & WSIs & Source & Added information & Comments\\
        \midrule
        TUPAC16 & 2016 & 821 & TCGA & Tumor and molecular proliferation scores & Ground truth is provided for 500 WSIs\\
        DRYAD & 2018 & 195 +40 +110 +239 & TCGA, CINJ, CWRU, HUP & Binary masks for annotated invasive regions in down-sized WSIs &  CINJ, CWRU, and HUP WSIs in full-size are not publicly available\\
        TIGER- WSIROIS & 2022 & 151 +26 +18 & TCGA, RUMC, JB & Region annotations on WSIs & All 195 WSIs have ROI polygons of different tissue regions and annotations of plasma and lymphocyte cells\\
        \bottomrule
    \end{tabular}
\end{table*}

\subsection*{Datasets description}

ACROBAT dataset~\cite{ACROBAT}: This dataset is part of the CHIME breast cancer study in Sweden, published in the AutomatiC Registration Of Breast cAncer Tissue (ACROBAT) challenge. ACROBAT entails 4212 WSIs from 1153 female primary breast cancer patients, where 1153 and 3059 images are H\&E and IHC stained, respectively. The slides are digitized using Hamamatsu NanoZoomer S360 and NanoZoomer XR scanners with 0.23 $\mu$m/pixel resolution. However, the published images are in TIFF format with 10$\times$ and lower resolutions (pixel size of 0.91 $\mu$m) to reduce the dataset size. In addition to the WSIs, the dataset includes annotations of landmark pairs between H\&E and IHC images for the validation (n=200) and test (n=606) sets.

ANHIR dataset~\cite{ANHIR}: This dataset is from the Automatic Non-rigid Histological Image Registration (ANHIR) challenge, which was part of the  IEEE International Symposium on Biomedical Imaging (ISBI) 2019. ANHIR contains whole-slide images of different types of tissue, including breast. Breast WSIs are stained with H\&E and IHC and scanned with Leica Biosystems Aperio AT2 with 40$\times$ magnitude and 0.253 µm/pixel resolution. Images are marked manually with landmarks with standard ImageJ structure and coordinate frame.

BACH~\cite{ICIAR2018}: The BreAst Cancer Histology images dataset is from the challenge held as part of the International Conference on Image Analysis and Recognition (ICIAR 2018). The dataset includes H\&E stained WSIs and patches. There are 400 patches with 2048$\times$1536 resolution, image-wise labeled in four different classes, along with annotations produced by two medical experts. BACH consists of 30 WSIs, acquired by Leica SCN400 scanner in SVS format, out of which 10 WSIs have coordinates of benign, in situ carcinoma, and invasive carcinoma regions, labeled pixel-wise by two pathologists.

BCNB~\cite{BCNB}: The Early Breast Cancer Core-Needle Biopsy WSI Dataset is the only publicly available dataset of breast histopathological WSIs from Asia. This dataset has 1058 WSIs from 1058 breast cancer patients in China. Images are scanned using an Iscan Coreo pathologic scanner, and tumor regions of each image are annotated by two pathologists. Furthermore, the clinical data, including the patient's age, tumor size, histology and molecular subtypes, number of lymph node metastases, and their status of HER2, ER, and PR, is made publicly available alongside the WSIs.

BRACS dataset~\cite{BRACS}: BReAst Carcinoma Subtyping dataset is collected at the Istituto Nazionale dei Tumori, Italy using an Aperio AT2 scanner at 0.25 $\mu$m/pixel for 40$\times$ resolution. BRACS contains 547 WSIs of 189 patients, labeled in 7 classes. Benign tumors are labeled Normal, pathological benign, and usual ductal hyperplasia. Atypia tumors are labeled flat epithelial atypia and atypical ductal hyperplasia, and malignant tumors have ductal carcinoma in situ and invasive carcinoma labels. In addition, 4539 regions of interest acquired from 387 WSIs are labeled and provided in .png files.

The Camelyon 16 and 17 datasets~\cite{Camelyon16, Camelyon17}: The Cancer Metastases in Lymph Nodes Challenge 2016 consists of 399 WSIs of H\&E stained lymph node sections collected in two centers in the Netherlands. Images are annotated with a binary label, and the ground truth for images containing metastases is available in WSI binary masks and plain text files in .xml format, providing the contour vertices of the metastases area. The dataset has 269 images in normal and metastasis classes for training and 130 WSIs for testing. The Camelyon 17 is the extended version of Camelyon 16 comprising 1399 unique H\&E stained WSIs, with an additional 1000 images added to the previous dataset. These 1000 WSIs are collected equally at five medical centers in the Netherlands, each providing 200 images from 40 patients (five slides per patient). In Camelyon 17, images of 100 patients are provided for training, and images of 100 other patients for testing. This dataset has detailed contours of metastasis boundaries on a lesion level for 50 WSIs and pN-stage labels for the patients in training data.

\begin{figure*}[!ht]
	\centering
	\includegraphics[scale=.8]{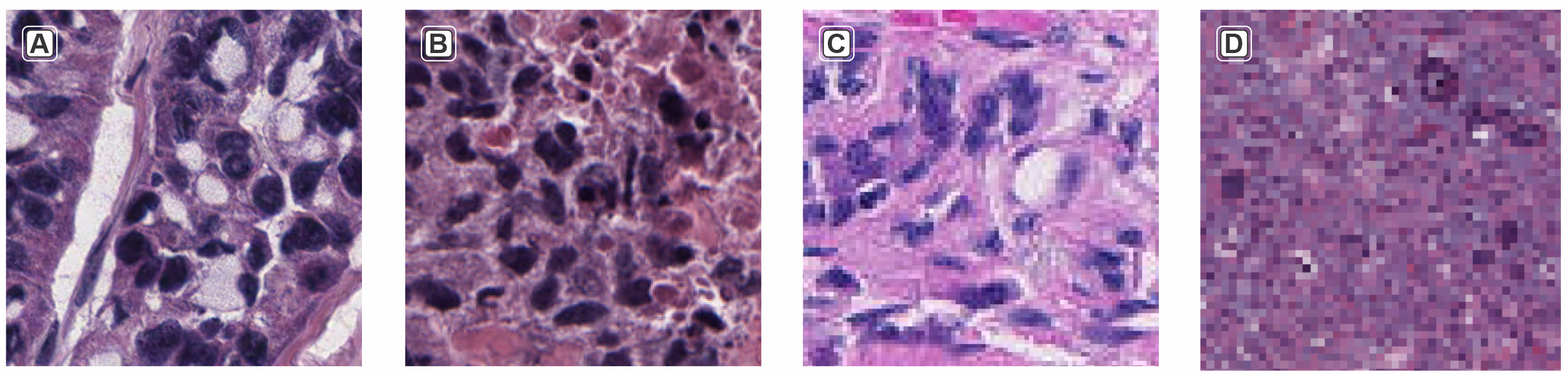}
	\caption{Comparative resolutions of four image tiles: (A) and (B) are two image tiles cropped from WSIs in the CPTAC-BRCA dataset with pixel widths of 0.25 $\mu$m and 0.5 $\mu$m, respectively. (C) shows an image tile from the ACROBAT dataset with a larger pixel width of 0.91 $\mu$m, resulting in reduced image resolution. (D) displays a tile from a down-sized WSI in the DRYAD dataset with an unknown pixel width, where the image quality is insufficient for cell detection purposes.}
	\label{FIG:3}
\end{figure*}

CPTAC-BRCA dataset~\cite{CPTAC-BRCA}: The Clinical Proteomic Tumor Analysis Consortium Breast Invasive Carcinoma Collection consists of 642 whole-slide images of 134 patients, scanned at 20$\times$ magnification. The published images have two different resolutions: 0.25 and 0.5 $\mu$m/pixel, which can be important when creating image tiles. (Fig.~\ref{FIG:3}). In addition to the slides, clinical, proteomics, and genomic data are available for researchers.

DRYAD dataset~\cite{DRYAD}: This dataset consists of 4 different cohorts: The Cancer Genome Atlas (TCGA), Cancer Institute of New Jersey (CINJ), Case Western Reserve University (CWRU), and Hospital at the University of Pennsylvania (HUP) each allotting 195, 40, 110 and 239 WSIs of breast tissue from ER+ patients. Slides are scanned by Aperio and Ventana whole-slide scanners at 40$\times$ magnification with 0.246 and 0.23 $\mu$m pixel width, respectively. Images in the published dataset are down-sized (32:1) WSIs with binary masks for annotated invasive regions. In~\cite{cruz2014automatic, celik2020automated, ektefaie2021integrative}, the authors did not mention any use of the DRYAD dataset, but they used the CINJ and HUP datasets that are included in DRYAD. Therefore, papers using these two datasets are included in our study, supposing that they have used part of the DRYAD dataset; as to our knowledge, the CINJ and HUP datasets are not separately available to the public in any database.

GTEx-Breast dataset~\cite{GTEx}: The Genotype-Tissue Expression (GTEx) project hosts gene expression levels of 44 human tissues. This project has published 894 breast tissue histology images, consisting of 306 and 588 WSIs of female and male breast tissues dissected from the central breast subareolar region of the right breast. The images are collected from different centers in the US and have short pathology notes. Additionally, GTEx provides an annotation file with detailed information about the samples.

HER2-Warwick dataset~\cite{HER2}: The data is part of the HER2 scoring contest organized by the University of Warwick, the University of Nottingham, and the Academic–Industrial Collaboration for Digital Pathology consortium. The dataset comprises 86 H\&E stained WSIs of invasive breast carcinomas acquired from 86 patients. IHC stained images, the ground truth data in the form of HER2 scores, and the percentage of cells with complete membrane staining are also provided in this dataset.

HEROHE dataset~\cite{HEROHE}: This dataset is presented in the HER2 on hematoxylin and eosin (HEROHE) challenge, aimed at predicting HER2 status in breast cancer by using only H\&E stained WSIs. This dataset entails 360 invasive breast cancer cases (144 HER2+ and 216 HER2-) for training and 150 cases (60 HER2+ and 90 HER2-) for testing. The WSIs in training and test sets are from different patients to maintain the independence between the two datasets. WSIs are scanned by 3D Histech Pannoramic 1000 in .mrxs format. Only a binary classification indicating positive or negative HER2 status is available for the HEROHE dataset, and the location of the invasive carcinoma is not annotated.

IMPRESS dataset~\cite{IMPRESS}: This dataset comprised 126 breast H\&E WSIs from 62 female patients with HER2-positive breast cancer and 64 female patients diagnosed with triple-negative breast cancer. All participants underwent neoadjuvant chemotherapy followed by surgical excision. In addition to the H\&E images, the dataset has IHC stained WSIs of the same slides and their corresponding scores. All the slides are scanned using a Hamamatsu scanner with 20$\times$ magnification. The IMPRESS dataset is published with clinical data (cohort metadata) for both patient groups, including patients' age and tumor size, as well as annotations for biomarkers such as PD-L1, CD-8, and CD-163.

Post-NAT-BRCA~\cite{Post-Nat-BRCA}: The Post neoadjuvant therapy (NAT) breast cancer dataset is from a cohort with residual invasive breast cancer following NAT. The dataset is composed of 96 WSIs from 54 patients. The slides were scanned by an Aperio scanner at 20$\times$ magnification at Sunnybrook Health Sciences Centre in Canada. Clinical data, including patients' age, ER, PR, and HER2 status, is also available with tumor cellularity and cell label annotations.

SLN-Breast~\cite{SLN-Breast}: The Breast Metastases to Axillary Lymph Nodes dataset consists of 130 H\&E WSIs of axillary lymph nodes from 78 patients, among them 36 WSIs have metastatic breast carcinoma. Slides were scanned with a Lecia Aperio scanner at 20$\times$ magnification at Memorial  Sloan Cancer Center in the US. Images are labeled in two classes, positive or negative breast cancer metastases.

TCGA-BRCA~\cite{TCGA-BRCA}: The Cancer Genome Atlas (TCGA) Breast Cancer study is an inclusive, experimental study of breast invasive carcinoma, coordinated and updated regularly by the US National Cancer Institute for research purposes. This dataset entails 3111 H\&E stained WSIs of breast cancer from 1086 Female and 12 male patients and is the largest publicly available dataset of breast histopathological WSIs. The TCGA-BRCA includes matched H\&E WSIs, gene expression data, and clinical information. We could not find any published region annotations for the WSIs, but there are external sources like cBioPortal that host comprehensive well-organized details of the patients in this dataset.

TIGER~\cite{TIGER}: this dataset is released in three formats as the training set for the Tumor InfiltratinG lymphocytes in breast cancER challenge. WSIROIS dataset has 195 whole-slide images from 195 patients with HER2+ and TNBC breast cancer. Images are collected from three sources: 151 WSIs of TNBC cases from the TCGA-BRCA, 26 WSIs from Radboud University Medical Center (RUMC) with both HER2+ and TNBC cases, and 18 WSIs of HER2+ and TNBC breast cancer cases from Jules Bordet Institut in Belgium. All the published images have 0.5 $\mu$m per pixel width and have annotated ROIs indicating seven different tissue regions, as well as 8$\times$8 $\mu$m\textsuperscript{2} bounding boxes indicating lymphocytes and plasma cells. The second dataset, called WSIBULK, consists of 93 WSIs from RUMC and JB with annotation of regions containing invasive tumor cells, and the third dataset, WSITILS, has only TIL values annotation of 82 WSIs without any manual region annotations. Images within WSIROIS, WSIBULK, and WSITILS are unique within each subset, with no duplications across the three subsets.

\subsection*{Datasets descriptive statistics}

\begin{figure*}[!ht]
    \centering
    \includegraphics[scale=.75]{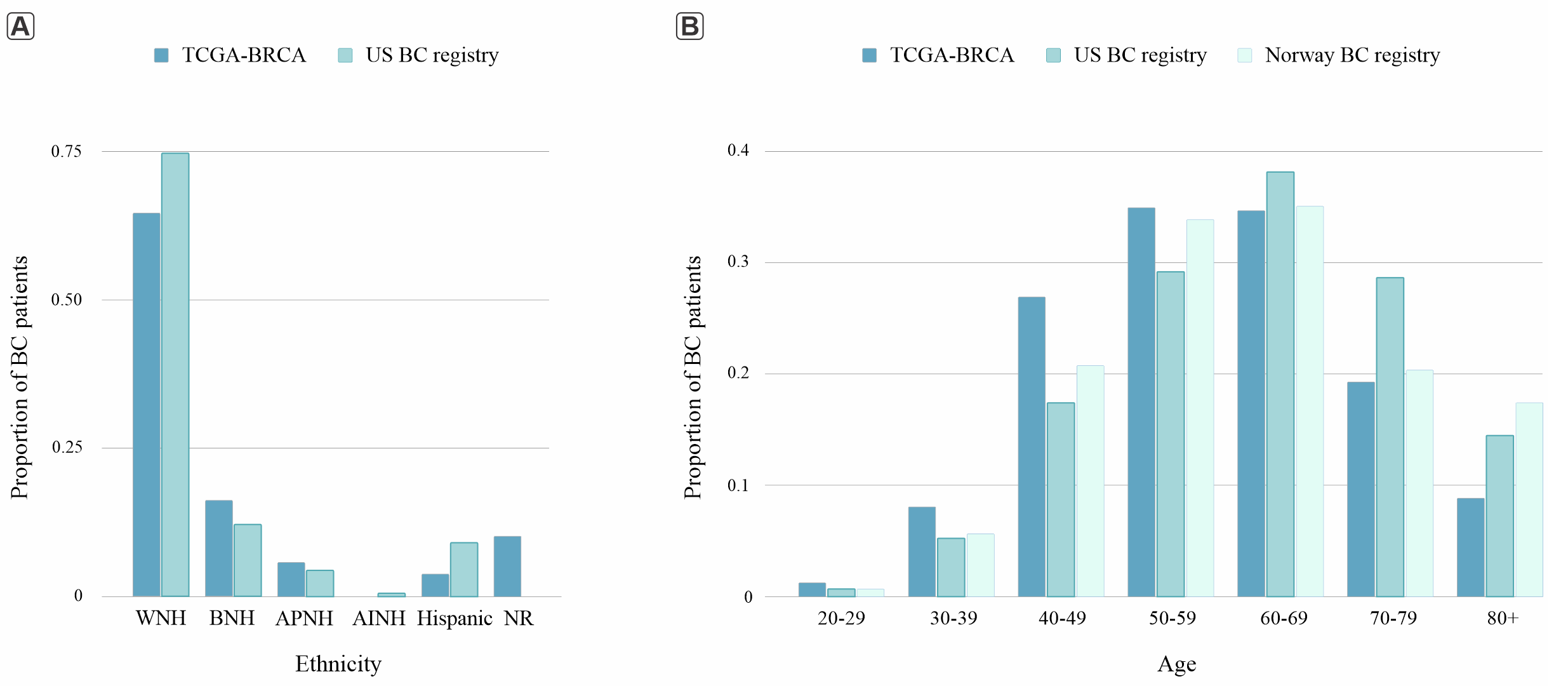}
    \caption{Publicly available breast H\&E WSI datasets usage across selected studies.
    (A) Usage frequency of breast public datasets in the included articles. The Post-Nat-BRCA, ACROBAT, and BCNB datasets were not used or mentioned in any reviewed articles. Note: In several included papers, multiple datasets are employed. (B) Share of active use of breast histopathology datasets for different tasks. The \emph{Other tasks} category comprises noise elimination, exploration of the tumor immune microenvironment, editing WSIs, feature engineering, investigation of bias in data, staining evaluation, histological grading, gene expression localization, crowdsourcing, obtaining tumor purity maps, scoring nucleoli, making image search and registration tools, and TIL assessment tools.}
    \label{FIG:4}
\end{figure*}

149~\cite{ahmed2021pmnet, amgad2019structured, anand2020deep, cruz2014automatic, aresta2019bach, bandi2018detection, bejnordi2017diagnostic, bokor2021weighted, brancati2022bracs, campanella2019clinical, ccelik2020resizing, celik2020automated, chaudhury2021novel, chen2019few, cho2021deep, ciga2021learning, ciga2022self, cruz2018high, cruz2017accurate, de2021machine, dhillon2020ebrecap, diao2021human, eddy2020cri, ektefaie2021integrative, elsharawy2020prognostic, elsharawy2021artificial, blanco2021medical, fernandez2016bagging, fischer2018sparse, ghazvinian2019impact, graham2019hover, guo2019fast, he2020integrating, hegde2019similar, howard2021impact, choudhary2019learning, jaber2020deep, jiao2021deep, kalra2020pan, kanavati2021breast, khened2021generalized, kim2020effectiveness, krithiga2020deep, kumar2020localization, kumar2017dataset, la2020detection, le2020utilizing, lee2021interactive, lei2020staincnns, levy2020spatial, li2021computer, li2021informed, li2021collagen, lin2019fast, Litjens2018camelyon, Liu2019detection, Lopez2021crowd, LU2022102298, Lu2021, LU20211032, WMi2021deep, Monjo2022, jimaging5030035, jimaging4020035, Munien2021, Munoz-Aguirre2020, Naik2020, Noorbakhsh2020, app10144728, Oner2021, OZTURK2019299, Pantanowitz2021, Park2021, Patil2020, PEREZBUENO2022102048, PEREZBUENO2021106453, Phan2021, Qu2021, RIASATIAN2021102032, Ruan2021, Runz2021, Saltz2018, Schmauch2020, SCHMITZ2021101996, SHAO2020101795, Sheikh2020cancer, SHI2020101624, SRINIDHI2022102256, Srivastava2018, Sui2021, Zhao2021, SUN201845, Sun2021, TELLEZ2019101544, Thagaard2021, Uchia2021, ValeSilva2021, Valieris2020, Valkonen2017, Venet2021, Vizcarra2019, LWang2021, WANG202289, YWang2021prediction, Wodzinski_2021, Wollmann2018AdversarialDA, WuC2019, Wulczyn2020, Xing2019, Xu2020, Xu2021predict, YANG2022333, Ye2019BreastCI, ZHANG2021275, ZhangW2020, ZHENG2019107, ZEISER2021115586, Bagchi2022, Chen2022, ChenS2023, Cong2022, Fassler2022, Foroughi2022, HuangZ2023, HuangJ2022, Jarkman2022, Jia2023, JIANG2023106883, XuJin2023, Lazard2022, LIU2022105569, LuW2022, Mondol2023, MouT2023, Sandarenu2022, SheikhTS2022, SunKai2023, TianJ2023, WangZ2022, WangRui2022, WuFei2022, Xingqui2022, Zheng2022Spatiality, Zheng2022Improving, SCHIRRIS2022102464, YiqingShen2022, Verdicchio2023, Dehkharghanian2023, Farahmand2022} out of the 176 included papers used public datasets of breast H\&E WSIs actively for different algorithm development purposes such as segmentation, classification, prognostic predictions, and color normalization of histology images. The remaining 27 articles~\cite{Cooper2018, DaiB2021, duggento2021deep, Gu2021Lesson, Hägele2020, hamidinekoo2018deep, JANSEN2020209, Lee2021cell, LiX2022, liew2021review, LITJENS201760, LopzM2016, Graziani2020CNN, Qaiser2018HER, Salvi2021, SCHNEIDER202280, SHAHID2019638, SOBHANI2021188520, SRINIDHI2021101813, STEINER2021188452, TRIPATHI2021101838, zeiser2021breast, Caldonazzi2023, Couture2022, KimInho2022, WuYawen2022, ZhaoYue2022} have reviewed or mentioned these datasets. These review papers are not included in the subsequent statistical analysis of datasets utilization in this paper. Fig.~\ref{FIG:4}A shows the frequency of active use of breast histopathology WSI datasets. Almost half of the studies (52\%) have used the TCGA-BRCA actively, which highlights the significance of this dataset in breast computational pathology and its value as a resource for future studies.

The TCGA-BRCA is the only dataset used for developing prediction models using breast WSIs, which might be explained by the fact that it has one of the largest cohorts among the publicly available datasets of breast WSIs, and it comes with clinical and genomics data. TCGA-BRCA has the largest contribution in the classification and \emph{Other tasks} categories and is the second most utilized dataset for detection/segmentation tasks, following closely behind the Camelyon dataset. Of Note, the Camelyon dataset stands out in the color normalization category as the sole dataset chosen for this particular task. (Fig.~\ref{FIG:4}B).

The available ground truth is a limiting factor in using public datasets of breast H\&E WSIs for computational pathology, especially when employing supervised algorithms for training the models. Only 43.3\% of WSIs have labels for breast cancer subtypes, 33.5\% of the images have annotations of regions of interest as ground truth, 11.4\% have binary labels of breast cancer metastasis, 5.1\% of images are provided by HER2 status labels or scores, and 6.7\% of the images do not have any annotations, which restrains the use of these datasets for specific tasks like classification of breast cancer subtypes. Nonetheless, the available WSIs can be utilized for training self-supervised and semi-supervised models.

\begin{figure*}[!ht]
    \centering
    \includegraphics[scale=.75]{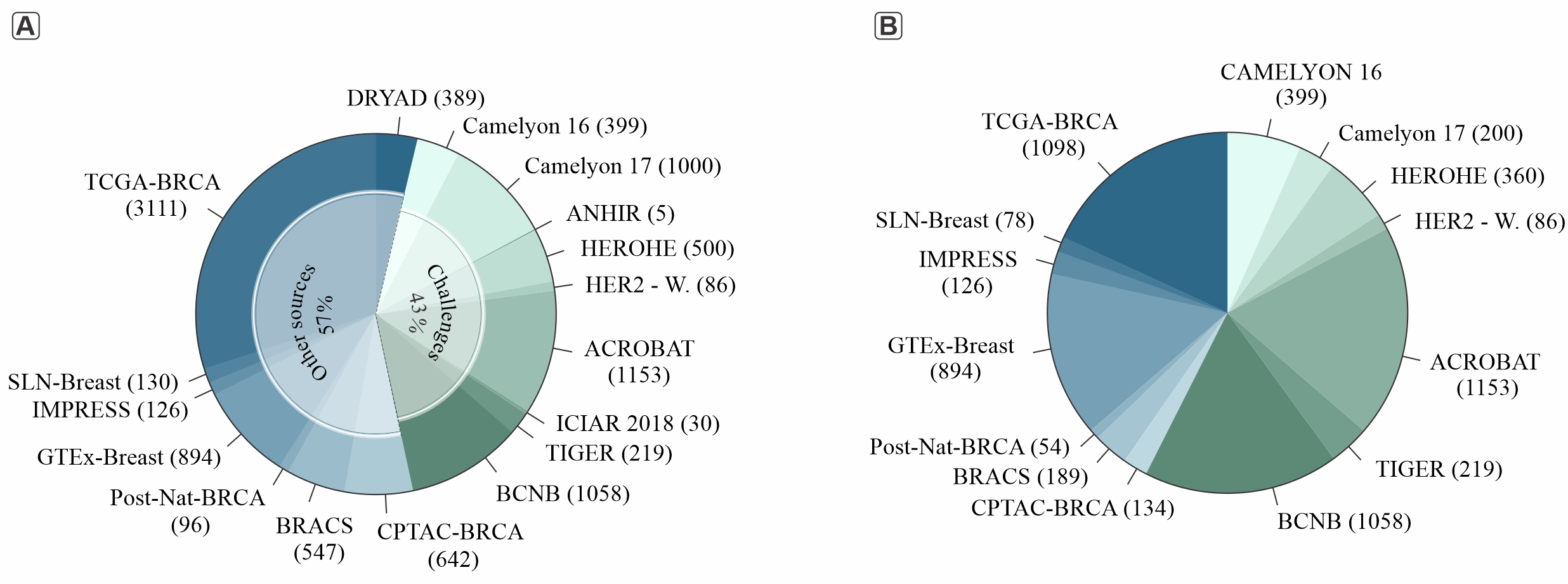}
    \caption{Distribution of images and patients in publicly available datasets of breast H\&E WSIs.
    (A) Number of WSIs in each dataset and proportion of image sources. The duplicate images in the DRYAD and TIGER datasets derived from the TCGA-BRCA are excluded. (B) Number of patients in each dataset. The number of duplicate patients in the TIGER dataset derived from the TCGA-BRCA (n=151) is not counted, and datasets that do not report the number of breast cancer cases are excluded.}
    \label{FIG:5}
\end{figure*}

Camelyon 16 and 17, HEROHE, ICIAR 2018, HER2-Warwick, ANHIR, ACROBAT, BCNB, and TIGER datasets are published in challenges and contests and comprise 43\% of the publicly available breast H\&E WSIs. The other eight datasets: TCGA-BRCA, SLN-Breast, Post-Nat-BRCA, BRACS, DRYAD, GTEx-Breast, IMPRESS, and CPTAC-BRCA are collected and made available for research purposes (Fig.~\ref{FIG:5}A). One intriguing aspect of the identified datasets is the number of patients included, which varies considerably between studies and may have important implications for the generalizability and reliability of trained models. The TCGA-BRCA is the largest open-access dataset of breast H\&E WSIs with 3311 images from 1098 patients (Fig.~\ref{FIG:5}B), which is extended regularly by adding new slides to the dataset.

\section*{Discussion}

\begin{figure*}[!ht]
    \centering
    \includegraphics[scale=.8]{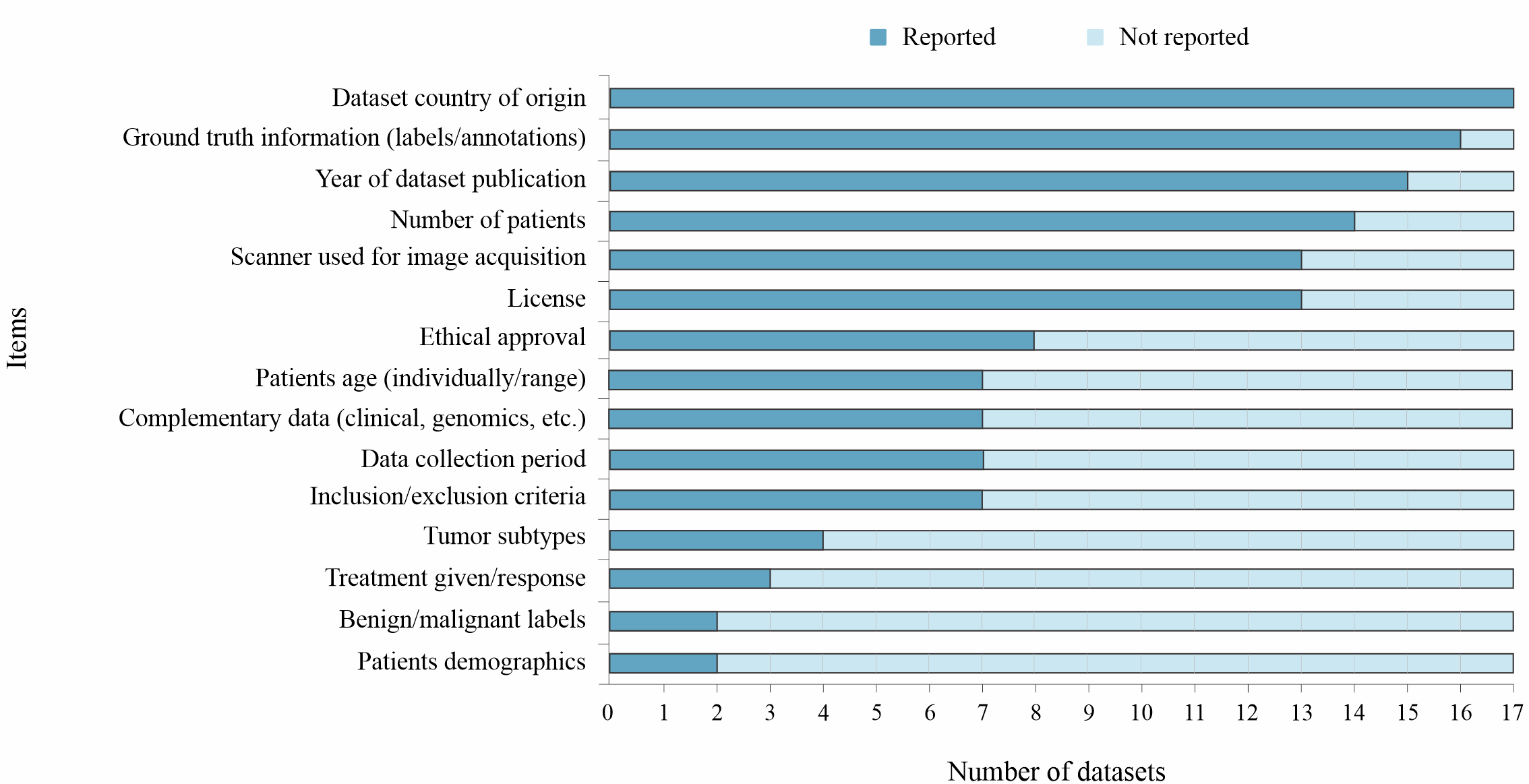}
    \caption{Report of public breast H\&E WSI datasets' characteristics and metadata.}
    \label{FIG:6}
\end{figure*}

The present study aimed to investigate the availability and suitability of open-access histopathology datasets for the development of deep learning algorithms in breast tissue analysis. In pursuit of this, we identified 17 publicly available datasets of breast H\&E WSIs that may appear to be a substantial amount for developing deep learning algorithms. However, it is important to note that the publicly available datasets of breast H\&E often lack detailed metadata descriptions (Fig.~\ref{FIG:6}). For instance, the number of patients is not reported in three datasets. Furthermore, the clinical data necessary for the development of prognostic tools are only available for six datasets: BCNB, Post-Nat-BRCA, TCGA-BRCA, GTEx-Breast, CPTAC-BRCA, and IMPRESS, with the latter four being collected in the US. TCGA-BRCA is the sole dataset published before 2021, which could explain its exclusive utilization for developing predictive models in the articles included in this study. None of the included papers or web pages hosting the breast H\&E WSI datasets provided an explicit statement of adherence to the FAIR principles, and the level of metadata and documentation provided by the dataset publishers varied. This variability in metadata and documentation could potentially affect the findability and reusability of the datasets, highlighting the need for improved adherence to the FAIR principles to enhance their accessibility and usability. Furthermore, inconsistencies in the data format and structure were found across the available datasets, which could limit the interoperability of the datasets.

\begin{figure*}[!ht]
    \centering
    \includegraphics[scale=.78]{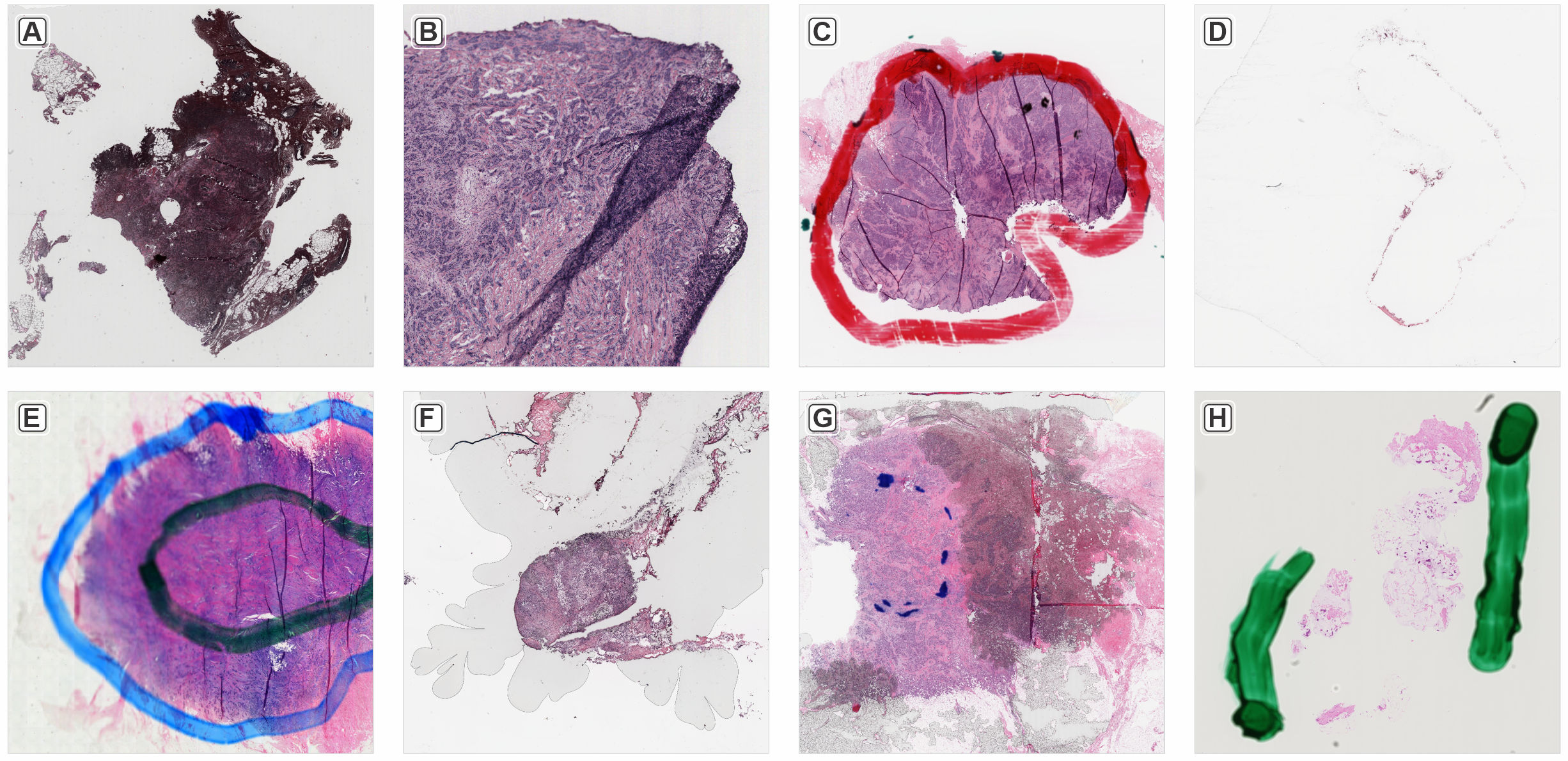}
    \caption{Examples of artifacts and quality issues in breast histopathological WSIs. (A) A WSI from the TCGA-BRCA with inconsistent staining. (B) Tissue folding in a WSI in the CPTAC-BRCA. (C) Pathologist's marker sign on a WSI from the TCGA-BRCA with several tissue foldings. (D) A slide from the CPTAC-BRCA exhibiting predominantly vacant areas. (E) A blurred WSI with marker signs from a private dataset. (F) Presence of air bubbles in a WSI of CPTAC-BRCA leading to focal image disruptions. (G) An image from the TCGA-BRCA with Marker signs, air bubbles, and inconsistent colors in some regions. (H) A WSI from the HER2-Warwick dataset showing marker signs on the slide.}
    \label{FIG:7}
\end{figure*}

The development of deep learning algorithms for breast computational pathology could be confounded by the quality of available WSIs. Variations in staining, tissue preparation, and image scanning processes can result in artifacts within whole slide images (WSIs), affecting the integrity of the data. Additionally, marker signs or annotations on the slides may inadvertently introduce noise or bias into the data. Therefore, accurate recognition and mitigation of such artifacts are vital to maintaining image-based analyses' fidelity. Fig.~\ref{FIG:7} shows examples of such artifacts and marker signs on the images.

\begin{figure*}[!ht]
    \centering
    \includegraphics[scale=0.8]{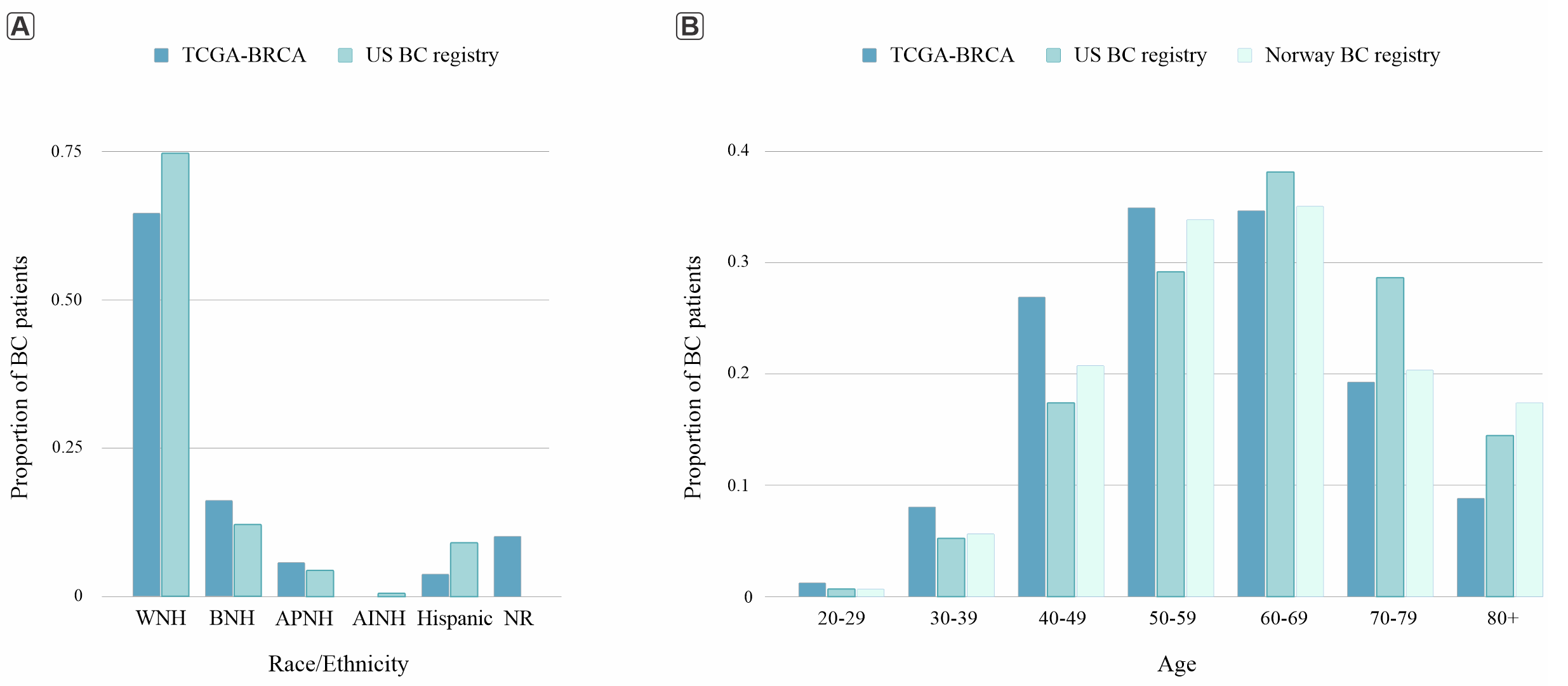}
    \caption{{Distribution of age and ethnicity among patients in the TCGA dataset.}
    (A) Distribution of breast cancer cases in different ethnicities. WNH=White Non-Hispanic. BNH=Black Non-Hispanic. APNH=Asian and Pacific Islander Non-Hispanic. AINH=American Indian and Alaska Native Non-Hispanic. NR=Not Reported. (B) Comparison of breast cancer cases in the TCGA-BRCA dataset, the US breast cancer registry~\cite{USCS} and the breast cancer registry of Norway~\cite{NORBCR}.}
    \label{FIG:8}
\end{figure*}

The WSIs of the TCGA-BRCA and the Camelyon datasets are widely employed in breast computational pathology. This widespread usage has implications for the generalizability of machine learning algorithms. As the TCGA dataset is collected in the United States, it may not be representative of the breast cancer population in other regions or countries. In addition, TCGA-BRCA has a high proportion of white women compared to American Indian and Hispanic patients (Fig.~\ref{FIG:8}A) and a high proportion of patients with infiltrating duct carcinoma (70\%). Additionally, this dataset includes a large percentage of samples from younger women, which may not represent the entire breast cancer population as the disease is more common in older women (Fig.~\ref{FIG:8}B). The composition of patient demographics is not reported in the Camelyon dataset, which limits the ability to analyze the potential impact of demographic diversity on the developed algorithms and findings.

The issue of biases is not exclusive to the TCGA-BRCA dataset; it has also been observed in other studies. A study on the representativeness of the TCGA bladder cancer cohort~\cite{SEILER2017} revealed biases in this dataset. The authors found that patients captured in the TCGA-BCa cohort demonstrate a higher risk disease profile compared to the reference cystectomy series, and consequently, their rates of overall survival and disease-specific survival are lower. Another study~\cite{Kim2019Racial} found that black Americans are not adequately represented in the majority of cancer cases within the TCGA datasets compared to clinical and mortality datasets. They also stated that Asian Americans are overrepresented in the TCGA dataset for most cancers. These biases are significant factors that should be acknowledged during the validation of computer-assisted tools, playing a vital role in maintaining the models' robustness, applicability, and transferability across different cohorts of breast cancer patients.

Deep learning models can benefit from external datasets to improve their ability to generalize to new data and enhance their performance on a specific task. Using external data to validate trained algorithms can help ensure their generalizability, identifying overfitting and assessing their performance across different datasets. Therefore, the real measure of a model's predictive ability lies in its performance on an independent dataset that was not employed in its initial development~\cite{Ivanescu2016}, as the performance of these models often diminishes when applied to a new cohort beyond the original development population~\cite{Collins2015}. Typically, there is a lack of external validation in the algorithms developed for breast computational pathology. Among the 149 non-review papers, only 41 incorporated multiple datasets; of those, 21 integrated private datasets alongside public ones for model development or validation. Notably, only 21 studies used an external validation/test set (Table~\ref{t3}), implying that the performance of other developed models could be subject to overestimation.

\begin{table}[!ht]
    \centering
    \setlength{\tabcolsep}{6pt} 
    \renewcommand{\arraystretch}{1.0} 
    \caption{Utilization of multiple breast H\&E WSI datasets in the included articles.}
    \label{t3}
    \small
    \begin{tabular}{p{0.8cm}>{\raggedright\arraybackslash}X p{6.1cm}>{\raggedright\arraybackslash}X p{5.2cm}>{\raggedright\arraybackslash}X p{3.1cm}}
        \toprule
        Study & Aim & Development dataset & External validation set \\
        \midrule
        ~\cite{ektefaie2021integrative} & BC Classification & TCGA, PD & DRYAD (HUP, CINJ) \\
        ~\cite{ahmed2021pmnet}  & Region detection and classification & ICIAR, DRYAD & TCGA\\
        ~\cite{anand2020deep}  & Detection and overexpression of HER2 & HER2-W. & TCGA \\
        ~\cite{bandi2018detection}  & Metastasis detection & Cam17 & PD \\
        ~\cite{cho2021deep} & Treatment prediction & TCGA & PD \\
        ~\cite{ciga2022self} & Segmentation and classification & TCGA, Cam16, Cam17, TUPAC, SLN & - \\
        ~\cite{elsharawy2020prognostic} & Scoring nucleoli in invasive BC & TCGA & PD \\
        ~\cite{blanco2021medical} & Editing WSIs with GANs & Cam16, Cam17 & - \\
        ~\cite{ghazvinian2019impact} & Metastasis detection & Cam16, Cam17 & - \\
        ~\cite{choudhary2019learning} & Staining evaluation & TCGA, ICIAR, TUPAC & Cam17 \\
        ~\cite{kanavati2021breast} & Classification of invasive ductal carcinoma & PD & TCGA \\
        ~\cite{khened2021generalized} & Segmentation of WSIs & Cam16, Cam17 & - \\
        ~\cite{li2021collagen} & Detection of fiber orientation disorder & TCGA & PD \\
        ~\cite{Lu2021} & Detection and classification & TCGA, CPTAC, Cam16, Cam17 & PD \\
        ~\cite{Naik2020} & Detection of estrogen receptor status & TCGA, PD & - \\
        ~\cite{app10144728} & Detection of HER2 status & HER2 & TCGA \\
        ~\cite{Pantanowitz2021} & Making an image search tool & TCGA, PD & - \\
        ~\cite{PEREZBUENO2022102048} & Color normalization and classification & Cam17 & Cam16 \\
        ~\cite{SCHMITZ2021101996} & Segmentation & Cam16, ICIAR & - \\
        ~\cite{Srivastava2018} &  Prediction of patient staging and node status & TCGA, TUPAC & - \\
        ~\cite{Zhao2021} & Detection of BC & Cam16 and 17 & - \\
        ~\cite{Sun2021} & Making a TILS assessment tool & PD & TCGA \\
        ~\cite{TELLEZ2019101544} & Color normalization & Cam16, TUPAC & - \\
        ~\cite{Thagaard2021} & Making a prognostic tool & TCGA, PD & - \\
        ~\cite{Valieris2020} & Prediction of DNA repair deficiency & TCGA & PD \\
        ~\cite{LWang2021} & Metastasis detection & Cam16, Cam17 & - \\
        ~\cite{WANG202289} & Histological grading & TCGA & PD \\
        ~\cite{YWang2021prediction} & Prediction of molecular phenotypes & TCGA & PD \\
        ~\cite{ZHENG2019107} & Color normalization & Cam16, Cam17 & - \\
        ~\cite{ZEISER2021115586} & Segmentation & TCGA, ICIAR, DRYAD & - \\
        ~\cite{Chen2022} & Segmentation & BRACS, PD & - \\
        ~\cite{ChenS2023} & Metastasis detection & TCGA, BRACS & - \\
        ~\cite{Cong2022} & Color normalization & Cam16, Cam17 & - \\
        ~\cite{Fassler2022} & Spatial characterization of TILs & TCGA, PD & - \\
        ~\cite{HuangZ2023} & Prediction of response to NAC & TCGA & IMPRESS \\
        ~\cite{Jarkman2022} & Testing the generalizability of a model & Cam16, Cam17, PD & - \\
        ~\cite{Lazard2022} & Prediction of DNA-repair deficiency & PD & TCGA \\
        ~\cite{LIU2022105569} & Prediction of BC recurrence & PD & TCGA \\
        ~\cite{LuW2022} & Detection of HER2 status & TCGA & HER2-W., PD \\
        ~\cite{TianJ2023} & BC classification & ICIAR, Cam16 & - \\
        ~\cite{Farahmand2022} & HER2 detection and response prediction & PD & TCGA \\
        \bottomrule
    \end{tabular}
    \begin{flushleft} studies utilizing non-breast datasets, Tissue Microarrays (TMAs), and image tiles along with the breast WSIs for training, validation, or as an external test set are excluded from this table. BC=Breast Cancer. PD=Private Data. Cam=Camelyon. NAC=Neoadjuvant Chemotherapy.
    \end{flushleft}
\end{table}

The potential benefits of using private datasets can make it well worth the effort to get access to such datasets. Incorporating private datasets in both the training and validation processes can improve machine learning models' diversity, representativeness, and overall performance. Private datasets can also contain unique or hard-to-obtain data that might not be available through public sources. For example, WSIs acquired from patients who have taken specific treatments like immunotherapy and the response to this treatment does not exist in any publicly available datasets of breast WSIs.

One potential limitation in this scoping review is the possibility of missing relevant datasets due to the search strategy or selection criteria used. To mitigate this limitation, a comprehensive and well-defined search strategy was developed to ensure that all relevant datasets were captured. Nevertheless, the Post-Nat-BRCA, ACROBAT, and BCNB datasets were not found in any of the papers identified in our selected literature databases, and we found them during the manual screening of research data repositories. Another limitation is the use of only English language search terms and inclusion criteria. This approach may have resulted in the exclusion of relevant datasets that were unavailable in English or primarily in other languages. Future systematic or scoping reviews of publicly available datasets of breast H\&E WSIs could benefit from broader search strategies that include searches in multiple languages. This would increase the likelihood of identifying relevant datasets that are not primarily in English and thus reduce potential language-related bias. However, the feasibility of such an approach will depend on the availability of resources, expertise in multiple languages, and the research question being addressed. Additionally, changes in the availability of datasets over time may further restrict the relevance and applicability of such reviews of publicly available datasets. The review should be conducted within a well-defined time frame to address this limitation, and the publication date of included studies should be clearly reported.

\section*{Conclusion}

In summary, our study examined the availability and suitability of publicly available datasets of H\&E stained histopathology WSIs that can be used in breast computational pathology. This data overview can save significant time and effort by providing a starting point without the need for setting up a data collection study.

Despite the significant number of WSIs, we found limitations in metadata descriptions, inadequate clinical data, and inconsistencies in the format and structure of datasets. Additionally, the presence of biases within widely used datasets, such as TCGA-BRCA, raises concerns regarding the generalizability of models. Therefore, it is crucial to improve adherence to FAIR principles, enhance metadata descriptions, and address biases. Moreover, incorporating diverse datasets, including private and external sources, promises to improve model performance and generalizability.

\section*{Appendix A. Supplementary data}

\paragraph*{Supplement 1}
\label{Supplement1}
{\bf The PRISMA-ScR checklist.}

\paragraph*{Supplement 2} 
\label{Supplement2}
{\bf Full search strings used in the data repositories.}

\paragraph*{Supplement 3} 
\label{Supplement3}
{\bf List of breast H\&E image tiles (patches) datasets.}

\paragraph*{Supplement 4} 
\label{Supplement4}
{\bf List of private datasets of breast H\&E whole slide images.}

\paragraph*{Supplement 5} 
\label{Supplement5}
{\bf Available clinical variables for breast H\&E WSI datasets}

\bibliographystyle{unsrtnat}

\bibliography{refs}

\end{document}